# Why Name Popularity is a Good Test of Historicity
## A Goodness-of-fit Test Analysis on Names in the Gospels and Acts

Luuk van de Weghe, Ph.D. ORCHID: 0000-0001-8710-503X
Independent scholar, Port Angeles, WA, 98362, USA
luukvandeweghe@gmail.com

Jason Wilson, Ph.D. ORCID: 0009-0003-7734-1110
Department of Math and Computer Science, Biola University, La Mirada, CA, 90639, USA
jason.wilson@biola.edu

**Abstract**
Are name statistics in the Gospels and Acts a good test of historicity? Kamil Gregor and Brian Blais, in a recent article in *The Journal for the Study of the Historical Jesus*, argue that the sample of name occurrences in the Gospels and Acts is too small to be determinative and that several statistical anomalies weigh against a positive verdict. Unfortunately, their conclusions result directly from improper testing and questionable data selection. Chi-squared goodness-of-fit testing establishes that name occurrences in the Gospels and Acts fit into their historical context at least as good as those in the works of Josephus. Additionally, they fit better than occurrences derived from ancient fictional sources and occurrences from modern, well-researched historical novels.

**Keywords**
Bauckham, Gospels and Acts, Gregor and Blais, historiography, onomastics, statistics, goodness-of-fit

**Acknowledgements**
*We would like to acknowledge the inaugural work of Biola University's Quantitative Consulting Center students Kharissa Kristi, Annika Miller, and Emily Sheng, as well as the valuable feedback of statisticians Don Lewis and Andrew Hartley. We also thank the anonymous reviewers of JSHJ, who provided additional feedback on several biblical and statistical portions of this paper. We also thank Lydia McGrew for critical input during the initial stages of this project and Willem Jan Blom for recommendations concerning data selection and methodology.*

Our topic has its roots in the publication of Tal Ilan's 2001 *Lexicon of Jewish Names in Late Antiquity (Part 1: Palestine 330 BCE – 200 CE).*[1] This lexicon, like a telephone book, catalogs the names of approximately 2500 persons into a single volume, giving us data related to name origin and name popularity statistics of Palestinian Jews living around the time of Jesus. For this enormous undertaking, and her three subsequent volumes, we owe her a debt of gratitude.[2]

New Testament scholar Richard Bauckham soon took advantage of this database to aid in the innovative study of names that resulted in his book, *Jesus and the Eyewitnesses*.[3] Herein he argued, for instance, that the names Levi (Mk 2:14) and Matthew (Mt. 9:9) unlikely refer to the same person because Ilan's database contains no example of a person bearing two popular Semitic names.[4] Bauckham also observed naming trends within the Synoptic Gospels, noting that Matthew and Luke sometimes drop names from Mark's text but do not add names to anonymous persons from Mark, supporting Markan priority and demonstrating a lack of creative tendency regarding names throughout the triple tradition.[5] Additionally, Bauckham discussed the historiographic aspects of Luke's preface and suggested that Luke used the literary device of *inclusio* to highlight his reliance on several named women, as well as Petrine testimony, to write his Gospel.[6] These and other arguments formed part of Bauckham's broader contention that, as in particular accounts from Plutarch's *Life of Julius Caesar* (*Caes.* 32.3–6; 46.2; 52) or Josephus' *Jewish War* (*War* 4.81–82; 7.398–99), the Gospel authors retained and emphasized certain names because these indicated the eyewitness sources behind portions of their narratives.[7] It is within this context that Bauckham also demonstrated, using Ilan's lexicon, that the name popularity statistics in GA generally conform to the population statistics attested to in Ilan's database.[8] Figure 1 below portrays Bauckham's insight in a graphical form, with the tops of the bars representing the percentages from Ilan's database.[9]

---

[1] Tal Ilan, *Lexicon of Jewish Names in Late Antiquity. Part I, Palestine 330 BCE–200 CE* (Tübingen: Mohr Siebeck, 2002).

[2] Her three other volumes are: Tal Ilan, *Lexicon of Jewish Names in Late Antiquity. Part III, The Western Diaspora 330 BCE–650 CE* (Tübingen: Mohr Siebeck, 2008); Tal Ilan, *Lexicon of Jewish Names in Late Antiquity. Part IV, The Eastern Diaspora 330 BCE–650 CE* (Tübingen: Mohr Siebeck, 2011); Tal Ilan, *Lexicon of Jewish Names in Late Antiquity. Part II, Palestine 200–650 CE (Tübingen:* Mohr Siebeck, 2012).

[3] Richard Bauckham, *Jesus and the Eyewitnesses: The Gospels as Eyewitness Testimony* (in this paper we cite the second edition; Grand Rapids: Eerdmans, 2nd. edn., 2017).

[4] Bauckham, *Jesus and the Eyewitness*, pp. 108–10; Bauckham argued that Judas son of James (Mk 3:18) and Thaddaeus (Lk. 6:16) could refer to the same person in light of common naming practices in Palestine (pp. 99–100).

[5] Bauckham, *Jesus and the Eyewitness*, pp. 40–44; Bauckham notes, on p. 44, that 'the practice of giving an invented name to a character unnamed in the canonical Gospels seems to have become increasingly popular from the fourth century on, but it is remarkable how few earlier examples are known.'

[6] On Luke's preface, see Bauckham, *Jesus and the Eyewitnesses*, pp. 116–24; for further discussion on the nature of Luke's preface, see Loveday Alexander, *The Preface to Luke's Gospel: Literary Convention and Social Context in Luke 1:1–4 and Acts 1:1* (SNTSMS; Cambridge: Cambridge University Press, 1993); Sean Adams, 'Luke's Preface (1.1–4) and its Relationship to Greek Historical Prefaces: A Response to Loveday Alexander', *JGRCJ* 3 (2006), pp. 170–91; J. J. Peters, 'Luke's Source Claims in the Context of Ancient Historiography', *JSHJ* 18.1 (2020), pp. 35–60. For Bauckham's discussion of *inclusio*, see *Jesus and the Eyewitnesses*, pp. 124–27; 132–45 (as it relates to the named women in Luke's Gospel, see pp. 129–31; as it relates to Simon Peter, see pp. 124–27); he responds to criticisms against the *inclusio* device on pp. 511–20, drawing on the implications of Mark writing 'contemporary historiography' (p. 511) and providing a historiographical parallel from Polybius' account of Scipio Africanus.

[7] Bauckham discusses the examples from Plutarch's *Life of Julius Caesar* and Josephus' *Jewish War*, as well as several others, only in his second edition of *Jesus and the Eyewitnesses*, pp. 524–34. Regarding Bauckham's main focus, note his comment from the foreword to the second edition: 'Perhaps one sentence may be cited as a summary: "the traditions were originated and formulated by named eyewitnesses, in whose name they were transmitted and who remained the living and active guarantors of the traditions."'

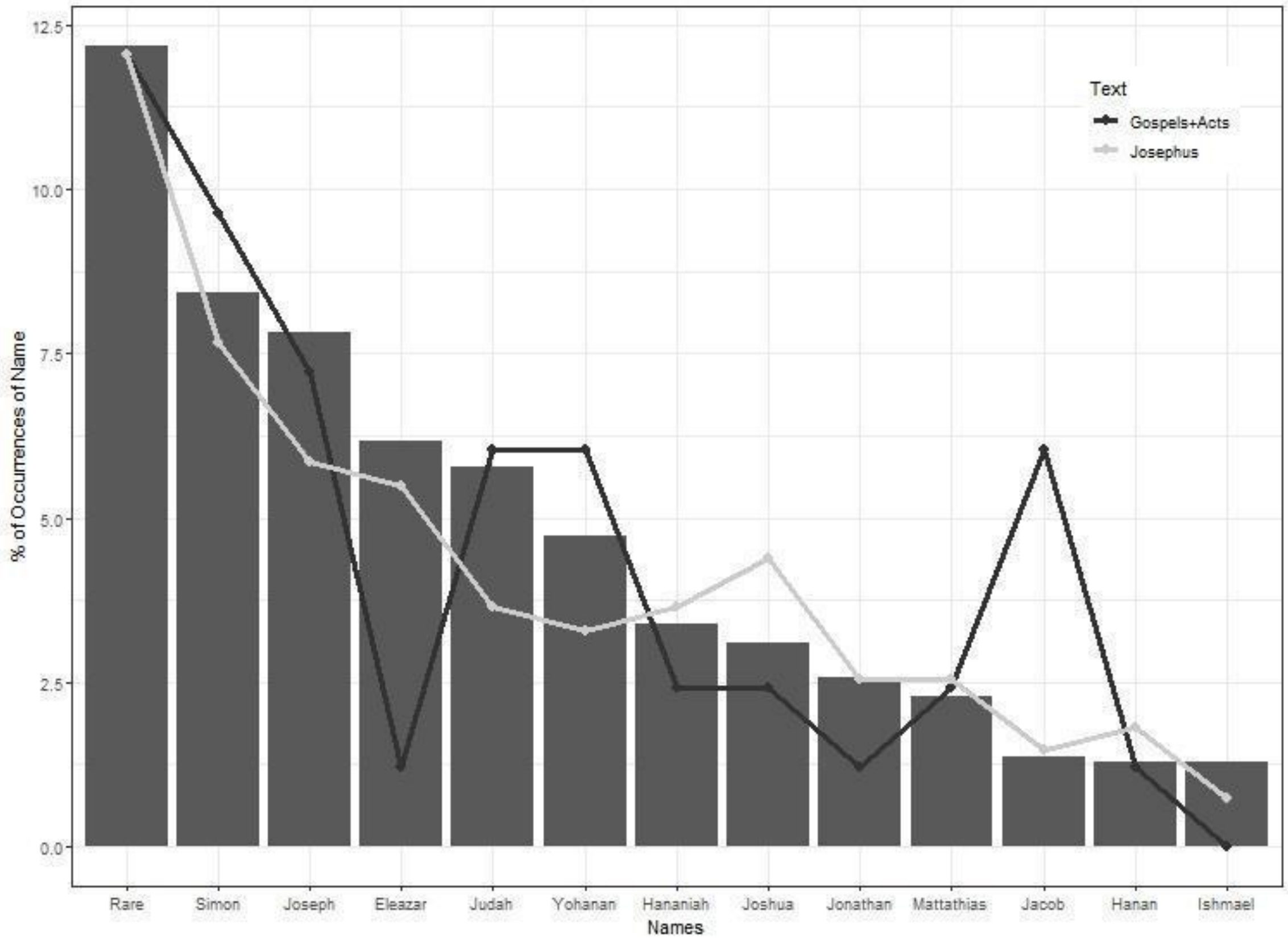

*Figure 1. Top 12 names %'s from Ilan-1 shown for Gospels+Acts and Josephus. The bar heights represent the height of the full Ilan-1 % of the name frequencies of the 12 most frequently occurring names. The %'s for Gospels+Acts, and Josephus, are superimposed with a dot. A connective line was added for ease of viewing. In addition, the % of rare names (=1 occurrence) was added.*

A question recently posed by Kamil Gregor and Brian Blais (GB) was whether this insight could withstand formal statistical analysis. They call the significance of Bauckham's observations into question while simultaneously furthering the conversation in this novel area of study.[10]

[8] Bauckham, *Jesus and the Eyewitnesses*, pp. 67–84.

[9] Although not perfect, the fit is striking. Eleazar (Lazarus) and Jacob (James) are off, although they fall within the confidence intervals, depending on how many name occurrences are considered in the Gospels and Acts sample. The data for Figure 1 is discussed in Section 2 below. Following Bauckham, this graph includes occurrences from Josephus and GA within the Ilan data, but in our statistical analysis we exclude them (see Section 3.1). We briefly revisit this graph in Section 2.

[10] Kamil Gregor and Brian Blais, 'Is Name Popularity a Good Test of Historicity?', *JSHJ* 21.3 (2023), pp. 171–202.

Few studies have otherwise sought to apply statistical analysis to onomastic data, and even here the aims and subject matter were quite dissimilar from Bauckham's project.[11] These parallel studies also lack in-depth discussions on statistical methodology. The scarcity of this literature highlights the extent to which Bauckham's approach – which had the potential of combining the hard and soft sciences in a single endeavor – was innovative. GB advance the conversation in several ways. First, they draw our attention to Ilan's complete 4-volume database in its current machine-readable form online.[12] Second, they rightly lament the lack of statistical rigor in Bauckham's analysis.[13] Finally, they point out several significant miscalculations by Bauckham and do well to constrain their analysis to Palestinian Jews from 4 BCE–73 CE in order to capture an onomastic snapshot of the time of Jesus.[14] They summarize their findings by stating that 'Bauckham's thesis offers no advantage in explaining the observed correspondence between name popularity in Gospels-Acts and in the contemporary Palestinian Jewish population over an alternative model of "anonymous community transmission"'.[15] We revisit this concept of 'anonymous community transmission' and GB's claim in our closing remarks to Section 3.

While advancing the conversation, critical missteps undermine GB's assessment of names in GA. For their analysis, they reduce the list of name occurrences in GA to 'contested names' (i.e., to only names of persons not corroborated by other primary sources) and then conclude that the sample is too small to draw statistically significant conclusions.[16] Due to their isolated focus on contested names and their improper testing methodology, GB fall short of providing the 'suitably robust' statistical analysis that their piece claims.[17] After presenting our

---

[11] See Dan Zao, 'Snack Names in China: Patterns, Types, and Preferences', *Names* 69.4 (2021), pp. 13–20 (16); Alessandra Minello, Gianpiero Dalla-Zuanna, and Guido Alfani, 'The Growing Number of Given Names as a Clue to the Beginning of the Demographic Transition in Europe', *Demographic Research* 45 (2021), pp. 187–220 (205); M. Depauw and W. Clarysse, 'How Christian was Fourth Century Egypt? Onomastic Perspectives on Conversion', *VC* 67.4 (2013), pp. 407–35 (424).

[12] For the full machine-readable database, see: https://github.com/hlapin/eRabbinica/tree/master/ilanNames (accessed January 9, 2024).

[13] Gregor and Blais, 'Name Popularity', p. 180; cf. Bauckham, *Jesus and the Eyewitnesses*, pp. 67–84.

[14] The period of 330 BCE–200 CE from Ilan, *Part I*, needed to be narrowed to achieve a 'snapshot impression of an onomastic situation', as Ilan calls it (Ilan, *Part I*, p. 50). See also, Luuk van de Weghe, 'Name Recall in the Synoptic Gospels', *NTS* 69.1 (2023), pp. 95–109 (96); the careful reader is immediately struck by the 371 name occurrences noted in this piece versus the 2181 noted by GB. The main reason for this discrepancy is not articulated in 'Name Popularity' and appears to be the difference between an inclusive versus exclusive approach in dating methodology. Take, for example, GB's time period of 4 BCE–73 CE. Many of Ilan's entries are dated broadly (e.g., –135 CE, –200 CE, –70 CE, etc.). Few are dated specifically (29 CE, 63 CE, etc.). In Van de Weghe, 'Name Recall', an exclusive approach is used, only counting a name occurrence if it could be dated specifically within 30 BCE–90 CE (pp. 105–06; note that p. 105 contains a typographical error and should read 371 names, not 391, per the table on p. 106). Gregor and Blais apparently take an inclusive approach, counting every name occurrence that cannot specifically be shown *not* to occur within their time window. In this article, we follow GB's timeframe and methodology.

[15] Gregor and Blais, 'Name Popularity', p. 171.

[16] This is because they narrowly conceive of Bauckham's project in terms of disproving the invention of every named person (Gregor and Blais, 'Name Popularity', pp. 185–97); therefore, their reasoning goes, if a person is already shown to be historical from other sources, it becomes irrelevant to their analysis. Our concerns with this position are further discussed in footnotes 17, 77, and 82 and in Sections 4.3 and 4.6 below.

[17] 'Name Popularity', p. 202. One immediate concern is the lack of definition around 'historicity'. As articulated in the previous footnote, GB do not specify but seem to imply a narrow definition, as if 'historicity' equals the

statistical analysis, we revisit these problems in Section 4 of this paper, where we provide a detailed critique of GB's article. Their study also contains several highly contentious speculations, but because our analysis broadly undermines the premises of almost every one of these there will be no need to address them in-depth.[18]

Below, we demonstrate that the Gospels and Acts, alongside Josephus, reflect a population of historical Palestinian Jewish names from 4 BCE–73 CE as opposed to a population of Diaspora males from the same period. Among other conclusions drawn from our analysis, we also demonstrate that name samples from the most robust historical novels available to us – novels that did indeed rely on Josephus, GA, and the Hebrew Bible for their naming practices – were not able to achieve naming statistics that significantly reflect Palestinian Jews from 4 BCE–73 CE. This latter analysis implicitly responds to a historically nuanced scenario envisioned by GB wherein the Gospel authors invented certain names but also relied on historical source substantiation of every contested name. Bauckham's view, as he admitted, never aimed for such a high bar (Bauckham, *Jesus and the Eyewitnesses*, p. 544). A widespread correspondence of names would naturally increase the likelihood of the historicity of each individual name, especially if we accept a model akin to Bauckham's wherein some of the name retention is explained by the informal citation of living informants; but a mere statistical analysis could not prove that every name in GA is historical, and a statistical analysis is not needed to disprove this point. For our purpose, we suggest that the name statistics in GA support 'historicity' to the extent that they demonstrate a level of information retention typical of ancient Greco-Roman historiography versus ahistorical and novelistic samples, and that the name popularity distribution of GA increases our confidence that we should understand these works in light of ancient historiographical practices. Our study improves upon and corroborates the claims in Van de Weghe, 'Name Recall', pp. 108–09, that a high level of accurate name retention most likely reflects historiographical practices, with potential implications for the type of oral tradition behind GA given the historical and religious context. This nuanced perspective moves us beyond merely 'proving' or 'disproving' every name occurrence. It invites further discussion about the implications for issues such as memory, oral tradition, and eyewitness testimony (per Bauckham's broader project) while allowing breathing room to consider the type of creativity allowable, and expected, in ancient historiography. For further discussions on reading GA in light of ancient historiographical practices, see footnotes 46 and 70 in this paper and the closing remarks to Section 3.

[18] For example, GB cite and discuss Dennis MacDonald's theory that Mark may have invented certain names to form literary pairs in his Gospel (e.g., one Simon 'Peter' who denies Jesus and another Simon 'of Cyrene' who carries his cross; Dennis Macdonald, *The Homeric Epics and the Gospel of Mark* (New Haven: Yale University Press), p. 23; *The Gospels and Homer: Imitations of Greek Epic in Mark and Luke-Acts* (Lanham: Rowman & Littlefield), p. 95, 108). While they concede that the theory is 'overly speculative' (Gregor and Blais, 'Name Popularity', p. 188), they suggest that it might explain the 'artificial inflation' of popular names in GA (p. 189); however, as demonstrated in Section 3.1 below, there is no artificial inflation of popular names in GA. Again, GB lean on the work of Steve Mason, *Josephus and the New Testament* (Peabody, MA: Hendrickson, 1992), pp. 185–229, and Richard Pervo, *Dating Acts: Between the Evangelists and the Apologists* (Santa Rosa: Polebridge Press; 2006), pp. 149–201, for the theory that the author of Luke-Acts relied on Josephus (Gregor and Blais, 'Name Popularity', p. 188), which, as the classicist Steve Reece observes, envisions Luke in the late first or early second century carelessly lifting details about the census of Quirinius, the rule of Lysanias, the revolts of Judas the Galilean, the *sicarii*, Theudas, and 'the Egyptian' from *Jewish War* and *Antiquities* while 'hopelessly distorting these details as he tries to incorporate them into his own narrative, and, finally, deliberately deceiving his readers in the preface of his work regarding the sources of his historical information (Luke 1:1–4) . . . it does not comport with what we witness in his work generally' (Steve Reece, *The Formal Education of the Author of Luke-Acts* (LNTS 669; London: T&T Clark, 2022), p. 79, n. 31). Regardless, even if we were to imagine such a scenario, this could not explain the onomastic data in Luke-Acts (see Section 3.2). Finally, GB's suggestion regarding the alleged greater propensity for fictionalizers to invent popular versus rare names (Gregor and Blais, 'Name Popularity', p. 200) is directly contradicted by the available evidence (see Section 3.2).

material found in Josephus.[19] In short, we show that name statistics in GA fit into their historical context well, and that they fit significantly better than fictitious samples.

## 1. Data

We incorporate seven sets of data into our analysis: two sets are reference distributions; five are test distributions. Reference distributions represent the real populations that our test distributions are compared against. For example, we will analyze name occurrences in GA (test distribution) and Josephus (test distribution) against name occurrences from two populations (reference distributions) in Ilan's database. The purpose is to see how a given test distribution fits, or does not fit, into a reference distribution.

Below, we highlight the seven datasets and detail any changes we made to GB's or Bauckham's data. These changes are strictly due to historical considerations; had we made no changes to Bauckham's list, our results would not be statistically different. Indeed, had we used the smaller GA sample from GB of only contested name occurrences, our results would not be statistically different. Whether we use our data, Bauckham's data, or GB's data, the differences in outcome are statistically indistinguishable.[20] Likewise, our interaction with Bauckham and GB cause us to consider the Gospels (along with Acts) under a single dataset, although we acknowledge that their shared canonicity is a later development. Had we considered Matthew, Mark, John, and Luke-Acts individually, the conclusions would have been the same.[21]

### *1.1 Reference Distributions*

We accept the Ilan reference distribution for Palestinian Jewish males (4 BCE–73 CE) that GB scraped from Hayim Lapin's machine-readable database. We independently scraped this data, but due to minor variations between our scraped dataset and GB's,[22] as well as their valuable inclusion of the Josephus data, we chose to use the GB dataset for Ilan Vol. 1 (Ilan-1) with slight modifications, resulting in 2185 name occurrences.[23] As a brief point of clarification, a name occurrence is the number of times a name is attached to a unique person. For example, Simon has eight occurrences in GA because it is attached to eight persons. But Peter also has one occurrence, even though that name is attached to someone already named Simon. This must be distinguished from the number of times a person's name is mentioned in a text, which is irrelevant to our exercise. A title like 'Simon Peter' would give two occurrences, even though it only refers to one person and regardless of how many times this person might be referenced in our source material. Finally, we only consider males in our analysis for the mere reason that antiquity's androcentric focus limits the available data for female name occurrences.[24]

---

[19] Gregor and Blais, 'Name Popularity', p. 188.

[20] The p-values for Bauckham's list of 79 names are 0.6106 (name frequency) and 0.0343 (name origin); the p-values for GB's list of 53 contested names are 0.2749 (name frequency) and 0.2794 (name origin). See Supplementary Materials. For a discussion of p-values, see Section 2.1 (cf. Table 3 below).

[21] The p-values for Matthew are 0.9396 (name frequency, 31 names) and 0.0965 (name origin); the p-values for Mark are 0.9844 (name frequency, 34 names) and 0.0197 (name origin); the p-values for Luke-Acts are 0.3998 (name frequency, 56 names) and 0.2185 (name origin); the p-values for John are 0.6322 (name frequency, 19 names) and 0.3940 (name origin). The names in Bauckham's Tables 1–4 in *Jesus and the Eyewitnesses* were used for these calculations. For a discussion of p-values, see Section 2.1 (cf. Table 3 below).

[22] GB has 2181 names. They described three different sets of names they excluded from the complete set of names from Ilan, *Part 1*: Nicknames, Fictitious names, and Other. Following GB, there were 58 Nicknames in Vol. 1 that occur only once in the entire Ilan database. This brought the list to 2488 names. The 'Other' exclusions included non-Palestinian people, such as those from Tarsus, and non-Jews, such as proselytes. Excluding all the Other and the Fictitious brought the number down to 2069 names. This is 112 less than GB. Since there were over 40 different categories in Other, some of which were a judgment call on whether they fit the non-Palestinian and non-Jew category, it is likely that the source of the discrepancy between the two datasets lies here. Such variation is not uncommon when scraping. Understandably, GB did not provide their list of exclusion categories, so it is neither possible nor worthwhile to further attempt to unify the two databases, and we opted to follow GB.

A second reference distribution covers the names of Western Diaspora Jewish males from Tal Ilan Vol. 3 (Ilan-3) within the same time parameters as GB's Ilan-1 (4 BCE–73 CE). We choose Ilan-3 as an alternative reference distribution to further situate Bauckham's claim that someone writing in the Diaspora would be exposed to significantly different Jewish name distributions than those of Palestine. The data from this reference distribution justifies Bauckham's claim, showing a significantly high frequency of Greek versus biblical names (discussed further in Section 3.1). Among the top ten names, only the name Simon from Ilan-1 is present. The top ten names are, in order, Shabtai, Dositheus, Ptolemaius, Alexander, Simon, Gaius, Theodotus, Julius, Theodorus, and Philon (cf. Figure 1).[25]

### *1.2 Test Distributions*

Aside from two reference distributions, we consider five test distributions. The first is the Gospels and Acts (GA). Like GB, we follow Bauckham's dataset, with three minor adjustments. Following Ilan, and in agreement with GB, we include the six Hellenists from Acts 6:5 (Stephen, Philip, Procorus, Nicanor, Timon, and Parmenas).[26] Additionally, we exclude Rufus and Alexander (Mark 15:21) because, being the sons of Simone of Cyrene, they should likely not be considered Palestinian. Lastly, we remove Bartimaeus from GA due to inconsistencies associated with leaving it in.[27] This results in 82 name occurrences for GA (79, per Bauckham,

---

[23] We added Qaifa (Caiaphas), which should have been included; although technically a family name, it evidently functions as a personal name in the relevant contexts. We also added the six Hellenists and removed Alexander, Rufus, and Bartimaeus for reasons discussed below. This resulted in 2185 name occurrences (2181, per GB, plus Qaifa, plus six Hellenists, minus Alexander, Rufus, and Bartimaeus).

[24] These clarifications are also discussed by GB ('Name Popularity', pp. 174–75) and in Van de Weghe, 'Name Recall', p. 99.

[25] All data and *R* code (www.r-project.org) used for the analysis, with some light explanation, is included in the Data and Supplementary Materials. https://doi.org/10.7910/DVN/N7API1

[26] GB state that there is no reason to assume the Hellenists, aside from Nicolas, are from the Diaspora ('Name Popularity', p. 185).

[27] We should not consider Bartimaeus as a name in GA because we already consider Timaeus as a name, which is how Ilan designates all the Bar-names – that is, by their patronyms. While Mark 10:45 contains the clause, 'Bartimaeus (which means "son of Timaeus")', this simply clarifies what is implicit in all the Bar-names. In all other cases, Bauckham agrees with Ilan's designations, and while Bartimaeus appears to function as a name in Mark 10:45, Bartholomew, which both Ilan and Bauckham designate by the patronym 'Ptolemy', functions the same way (Mk 3:18). If we catalog this name under the form 'Bartimaeus', we should also catalog Barsabbas, Bartholomew, etc., under the forms with their Bar- elements included rather than under the patronyms they derive from, but doing

plus six, minus three). A second test distribution we consider are the name occurrences in Josephus; we follow the helpful sample scraped by GB, with the addition of Qaifa.

Our third test distribution is the complete set of fictitious occurrences found in Ilan Vol. 1 (Ilan-1F). This sample derives from all name occurrences from 4 BCE–73 CE belonging to Palestinian Jewish males that Ilan deemed fictitious. This pool springs from noncanonical Christian literature as well as rabbinic material only loosely tied to authentic traditions.[28] Unlike the hypothetical uniform distribution considered by GB (discussed below), this sample derives from concrete data. We want to test how well the statistics in Ilan-1F fit, or do not fit, into the Ilan-1 reference distribution. This will give us an impression of how ancient fictionalizers performed in attempting to create authenticity in their naming practices for Palestinian Jewish males around the time of Jesus. Since this sample lumps many smaller discrete samples together, it allows data from fictitious sources with scarce onomastic data to be considered.

Our fourth test distribution is a complete sample of Palestinian Jewish male name occurrences from the novels *Ben Hur* and De Wohl's *The Spear*. Here we want to determine how well-researched historical novels perform in acquiring onomastic verisimilitude. This is significant, because GB imagined a scenario wherein GA could achieve an appropriate reflection of Ilan-1 by inventing names inspired by the works of Josephus as well as incorporating actual names of historical persons from various traditional materials.[29]

Louis De Wohl's *The Spear* is a meticulously researched historical novel set around the time of Jesus, incorporating many historical and fictional Palestinian Jewish characters.[30] Lewis Wallace's 1880 *Ben Hur*, while containing fewer Jewish names than *The Spear*, is a more familiar alternative. We therefore analyze all male name occurrences for both. Each work shows a wealth of historical data and research as well as unambiguous signs of dependence on Josephus and on the Gospels and Acts. Over 35 percent of their name occurrences appear to be directly influenced by these two sources.[31] Regarding Second Temple Judaism, Lew Wallace, the author of *Ben Hur*, claims to have visited the Library of Congress in 1873, researching, 'everything on the shelves relating to the Jews'.[32] These historical novels provide a good test case for what a well-informed inventor of Palestinian names might achieve in terms of appropriate naming patterns.

Finally, we consider a test distribution consisting of a sample of 52 occurrences from a uniform distribution. GB suggested an additional scenario of an inventor randomly selecting names from a uniform distribution of the 457 Palestinian names that made up their Ilan-1

---

so would render all these names rare and go contrary to the practice followed by Ilan everywhere and by Bauckham everywhere but in this instance.

[28] It also includes occurrences from several inscriptions and characters from stories such as Jesus' Lazarus from Lk. 16:20. Regarding post-Talmudic literature, Ilan writes, 'some of these stories may be based on authentic traditions, but with the passage of time and the literary nature of these compositions this is not very likely' (*Part I*, p. 48).

[29] Gregor and Blais, 'Name Popularity', pp. 188–89.

[30] We thank Lydia McGrew for bringing De Wohl's *The Spear* to our attention for this analysis.

[31] 32 out of 86 occurrences are persons specifically mentioned in GA or Josephus: 37.31%. In *Ben Hur* alone, the percentage is even higher (19 out of 32 occurrences: 59.38%).

[32] Lew Wallace, *Lew Wallace: An Autobiography*, (Vol. 2; Harper & Brothers: London, 1906), p. 891.

reference distribution; they argued that their sample of contested GA occurrences (53) was too small to significantly distinguish it from a random sample from Ilan-1.[33] We want to determine if this claim is true using our methodology, even though we take issue with their scenario (see Section 4.3; 4.6).

### *1.3 Data Pre-Processing*

Unfortunately, GB ignore data from names that have low occurrences in the GA test distribution: what they call 'white noise'.[34] This common problem in data analysis is easily overcome by the widely used pre-processing procedure called data binning. Data binning takes data from an intricate set and categorizes it into discrete bins for the purpose of simplification, noise reduction, and enhanced analysis and modeling.[35] This technique allows us to lump even low-frequency occurrences from test distributions into categories.

Consider, for example, the data observed in the beginning of this paper in Figure 1. While Figure 1 gives great details of the most popular and the rarest name occurrences, it provides zero detail about more than half of the names in GA; had we not binned the rare names for this graph, it would represent less than half of the names in GA. One might compare this to performing an in-depth analysis on voting preferences among Gen Z and Millennials in order to draw conclusions about the voting trends in all demographics. Clearly, focusing only on those groups will provide more detail about how particular age groups are voting, but it is not the proper procedure for making decisions regarding the demographics of an entire population.

Especially with smaller sample sizes, as those we are dealing with, some mismatches are inevitable (e.g., Eleazar and Jacob in Figure 1). But Gregor and Blais claim that 'even if there were only one Simon in Gospels-Acts, this would not fit the distribution of name popularity in the contemporary population statistically significantly worse than what we actually observe in Gospels-Acts! The same would be true even if *every* name among contested Gospels-Acts characters appeared only once.'[36] But that is only true under their model, which insufficiently weighs the cumulative force that each low occurrence would have within a model like the one we use in this paper. In short, binning allows us to consider more subtle features of name distributions than merely the most popular names (See further discussion below, Section 3.2).

For our binning, we considered the following features: objectivity, consistency, and robustness. Regarding objectivity, we binned all data *relative to each reference distribution*; it was the reference distribution rather than the test distribution that determined the sizes of bins. This generated consistency in how we analyzed each test distribution. We utilized a method called equal frequency binning, meaning that we binned name frequencies into sets with approximately the same amount of name occurrences from our reference distributions. Equal frequency binning minimizes the distance between the test and reference distributions, which is the mathematically least favorable approach for our purposes.[37] Because our smallest possible bin (one occurrence) encompassed approximately 1/6 of occurrences, the result was six bins.[38] This generated consistency across our tests of the popularity data. In fact, our conclusions are robust: experimenting with different bin sizes and methods yielded only the same results. Our procedure will become clearer in Section 3.

---

[33] The discrepancy of our 52 names versus their 53 names results from the removal of Bartimaeus from our distribution. For their discussion of this scenario, see Gregor and Blais, 'Name Popularity', pp. 191–95.

[34] Gregor and Blais, 'Name Popularity', p. 191; also, see our comments in Section 4 below.

[35] It is standard statistical practice. See a formal discussion of this procedure in the reference text by Alan Agresti, *Categorical Data Analysis* (Hoboken, NJ: Wiley & Sons, 2002), pp. 174–77.

[36] Gregor and Blais, 'Name Popularity', p. 194.

Additionally, we improved on GB's analysis by considering name origin statistics in addition to name frequency statistics. Tal Ilan categorized name occurrences into eight categories according to name origin: Biblical, Greek, Latin, Persian, Egyptian, Arabian, Semitic-Hebrew, and Semitic-Greek. A strong case could be made for combining the latter two categories into a single category (Semitic), but because it does not significantly impact our analysis we opted to follow Ilan's designations. We analyze these categories as an independent exercise from our name frequency analysis.

## 2. Method

In this paper, our research hypothesis is that the name occurrences in the Gospels and Acts are at least as historical as those from Josephus. In the preceding, we have considered literary and historical data. In what follows, we consider the statistical evidence. In particular, we define the statistical goodness-of-fit test, explain how it is used, and formally perform eighteen goodness-of-fit tests of the test distributions against the reference distributions.

### *2.1 Goodness-of-fit Tests*

For the analysis, we chose to use the most common and widely used statistical test of fit between a categorical test distribution and a categorical reference distribution: the chi-squared goodness-of-fit test.[39] For example, suppose that someone wanted to determine whether a die was fair or not. The reference distribution would reflect a probability of 1/6 (i.e. 'fair') for rolling a 1, 2, 3, 4, 5, or 6. Suppose we rolled the die 60 times and obtained the sample observed in Table 1.[40]

---

[37] 'The least upper bound of the "distances" of such alternative distributions from the null hypothesis distribution can evidently be minimized by making the [bin] probabilities under the null hypothesis equal to each other' (p. 306). This is from the seminal paper on this topic: H. B. Mann and A. Wald, 'On the Choice of the Number of Class Intervals in the Application of the Chi Square Test', *Ann. Math. Stat*. 13.3 (1942), pp. 306–17. Because of the uneven name frequencies, exact equal frequency bins were not possible. We used the following method to determine unique bins: taking the minimum root mean square error for different distributions of the name frequencies and selecting the one with the smallest root mean square error.

[38] It was a little less for Ilan-1 and a little more for Ilan-3; see the gray bars of Figure 2. When selecting the number of bins, two competing objectives must be balanced: more bins for more granularity versus fewer bins to have data and meet test conditions (no bin have fewer than one expected observations and fewer than 20% of bins have below five expected observations). Six bins met the assumptions in all but one formal case (where five bins were used, see Supplementary Materials).

[39] Alan Agresti, *Categorical Data Analysis* (Hoboken, NJ: Wiley & Sons, 2002), pp. 22–26. One alternative considered was a chi-square test of independence, which would treat both the test and reference distributions as samples (the goodness-of-fit test treats the reference distribution as the ground truth). For robustness, we ran the chi-square tests of independence. The p-values were quite close to the goodness-of-fit p-values shown in Table 3. They were usually slightly larger than those of the goodness-of-fit, but not always, and in no cases had any substantial divergence which would alter our conclusions. They are shown in the Supplementary Materials.

| **Face** | **1** | **2** | **3** | **4** | **5** | **6** | | **Total** |
|---|---|---|---|---|---|---|---|---|
| **Reference Distribution** | 1/6 | 1/6 | 1/6 | 1/6 | 1/6 | 1/6 | | **6/6=1** |
| **Observed** | 5 | 8 | 9 | 8 | 10 | 20 | | **60** |
| **Expected** | 10 | 10 | 10 | 10 | 10 | 10 | | **60** |
| **Difference** | -5 | -2 | -1 | -2 | 0 | 10 | | **0** |

*Table 1. Example of testing whether a 6-sided die is fair.*

Based on the observed values, do you think the die is fair, or not? The die being 'fair' means 'matches the reference distribution' and is called the **null hypothesis**. The die being 'not fair' means 'does not match the reference distribution' and is called the **alternative hypothesis**. These technical terms will be needed in Section 4.[41] The test works by using the reference distribution to calculate the expected value, which in this case is (1/6)*60 = 10 rolls expected for each of 1, 2, 3, 4, 5, and 6. The observed values are compared against the expected values and the probability value (*p-value*) and *power* are calculated.

The *p-value* is defined as the probability that the differences between the observed and expected values would be as great as the differences we observed, or greater, if the data truly came from the reference distribution. In this case, the *p-value = 0.0199*, meaning that there is about a *2%* probability that we would observe differences this large – merely five 1's but double the expected 6's – if the die is truly fair. In other words, this means that there is about a 2% chance of concluding the die is not fair when actually it is.

Does this prove that the die is fair or not fair? No, it only offers a probabilistic answer. In current scientific practice, p-values are compared against an arbitrary pre-selected benchmark, often *0.05* or *5%*. If the p-value falls below the benchmark, it is concluded that the sample distribution does not come from the reference distribution and the test is said to be **statistically significant.**[42] In this example we would have observed *p-value = 0.02 < 0.05*, meaning we conclude the die is not fair. The benchmark is the largest probability of making a false positive error that the researcher is willing to accept.

[40] This particular example is taken from Wikipedia's Pearson's chi-squared test page, https://en.wikipedia.org/wiki/Pearson%27s_chi-squared_test. Accessed January 12, 2024.
[41] For an introduction to these concepts, see Diez, David M., Christopher D. Barr, Mine Cetinkaya-Rundel; *OpenIntro Statistics,* 4th Ed. (2015); OpenIntro; https://open.umn.edu/opentextbooks/textbooks/60. Accessed May 14, 2024.
[42] GB also use this technical phrase, as will be mentioned below. 'Statistical significance' and p-values have been the subject of recent debate. In 2019, The American Statistical Association, the largest body of professional statisticians in the world, devoted an entire issue of one of their flagship publications, *The American Statistician* (Volume 73, Supplement 1), to the topic with a whopping 43 articles following the 19 page editorial. Their president also convened a task force which created a document which concludes, 'In summary, *p*-values and significance tests,

Suppose we go with $p\text{-}value = 0.02 < 0.05$ and conclude the die is not fair, i.e. the probabilities are not {.167, .167, .167, .167, .167, .167}. What is the chance that we are right? That is the power: *power* is the probability that the statistical test detects the difference between the test and reference distributions when there really is a difference. If the die is slightly biased, e.g.{.133, .133, .133, .200, .200, .200}, then $power = 0.189$, meaning there is only about a *19%* chance that, if the die is biased that way, we will correctly detect it! By contrast, if the die is heavily biased, e.g. {.083, .083, .083, .25, .25, .25}, then $power = 0.952$, meaning there is about a 95% chance that, if the die is biased that way, we will correctly detect it. Notice that the greater the distance of the alternative hypothesis from the null hypothesis/reference distribution, the greater the power, which is the ability to detect the difference.

Accompanying the statistical conclusions of any properly conducted statistical study should be all additional available scientific evidence and reasoning which are combined using logic to draw conclusions about the research hypothesis(es). In non-experimental studies, such as this one, p-values alone should not be used to draw definitive conclusions. That is the process we will use in this paper.

There is one additional remark needed regarding the benchmark against which we will compare our *p-values*. When multiple tests are performed, as in our case, the so-called Multiple Testing Problem occurs. This problem is best captured in a cartoon where scientists perform tests for whether colored jelly beans cause acne using a benchmark of *5%.*[43] They test *20* different jelly bean colors, and none are statistically significant – except the green one. The newspaper headline reads, 'Green Jelly Beans Linked to Acne!' Since *20* tests are performed, on average $1/20 = 5\%$ false positives are expected. Accepting the 'green jelly beans linked to acne' conclusion is a result of the multiple testing problem – if you run enough tests even when there is nothing, you will eventually 'find' something by chance. There are a variety of procedures to adjust the benchmark to account for this. We will use the Bonferroni correction[44] because it is the simplest, it is widely used, and using more sophisticated methods would not change our conclusions. The Bonferroni adjustment is applied by dividing the benchmark by the number of tests. In our case, we will have eighteen tests, therefore we will use the Bonferroni-adjusted benchmark of $0.05/18 = 0.0028$.[45]

---

when properly applied and interpreted, increase the rigor of the conclusions drawn from data…. Although all scientific methods have limitations, the proper application of statistical methods is essential for interpreting the results of data analyses and enhancing the replicability of scientific results' (https://magazine.amstat.org/blog/2021/08/01/task-force-statement-p-value/, accessed May 13, 2024).

[43] XKCD, 'Significant', https://xkcd.com/882/. Accessed on January 12, 2024.

[44] See R. G. Miller Jr., *Simultaneous Statistical Inference* (New York: Springer-Verlag, 1991).

[45] In this paper, we are not using 0.0028 as a magic threshold upon which all p-values below it are statistically significant and all above are not. May it never be! Indeed, as one reader correctly pointed out, we could select different numbers of tests to adjust this number. We merely offer it as an interpretive guideline on how to read the p-values shown below. The specific number is not important. We could have done as few as four tests (GA popularity and origin versus Ilan-1 and Ilan-3) and many more than eighteen. This could give Bonferroni-adjusted benchmarks

### *2.2 Design of Tests and Hypothesis*

Recall that we have five test distributions (observed samples) of popularity/frequency data for our analysis: (1) GA, which is the focus of our study, (2) the writings of Josephus, reflecting products of historiography,[46] (3) historical novels, (4) the fictitious occurrences from Ilan-1 (Ilan-1f), and (5) GB's uniform distribution. We chose two different reference distributions: Ilan Vol. 1 Palestinian Jews (Ilan-1) and Ilan Vol. 3 Diaspora Jews (Ilan-3).

| Reference Distribution | Variable | Gospels & Acts | Josephus | Novels | Ilan-1F | Uniform |
|---|---|---|---|---|---|---|
| Ilan-1 Palestinian Male Jews | Name Frequency | Fit | Fit | Not fit | Not fit | Not fit |
| | Name Origin | Fit | Fit | NA[47] | Not fit | Not fit |
| Ilan-3 Diaspora Male Jews | Name Frequency | Not fit | Not fit | Not fit | Not fit | Not fit |
| | Name Origin | Not fit | Not fit | NA | Not fit | Not fit |

*Table 2. The eighteen designed chi-square goodness-of-fit tests and our hypotheses.*

---

in the range of 0.05/4 = 0.0125 to 0.05/100 = 0.0005. The point is that 'small' p-values indicate statistical significance and 'large' do not, and 0.0028 (or 0.0125 to 0.0005) will be a guideline and not a rule.

[46] This genre classification does not make a composition immune from exaggeration and error. Colin Hemer remarks:

> Josephus, while an invaluable witness to matters within his experience, is prone to sensationalize and exaggerate . . . Josephus dwells on the horrors of the famine in doomed Jerusalem (War 5.10.3.429–38) or the grotesque fates of refugees (War 5.12.4.548–52), or a mother eating her child, a scene complete with speeches (War 6.3.3.199–4.213). This problem is especially apparent in the treatment of numbers. In Acts, 4,000 men follow the Egyptian bandit into the desert, and in Josephus, 30,000. Josephus has "not less than three million Jews" in an anti-Roman demonstration; can we accept such a figure?

(*The Book of Acts in the Setting of Hellenistic History* (ed. C. H. Gempf; Winona Lake, IN: Eisenbrauns, 1990), pp. 97–98). See also, Tal Ilan and Jonathan J. Price, 'Seven Onomastic Problems in Josephus' "Bellum Judaicum"', *JQR* 84.2/3 (1993), pp. 189–208. For a recent discussion on Josephus' rhetorical embellishments, his internal contradictions, and the difficulties with reconstructing a history 'behind Josephus' when Josephus is our only source, see Steven Mason, *Josephus, Judea, and Christian Origins: Methods and Categories* (Peabody, MA: Hendrickson Publishers, 2009), pp. 103–37. For a comparison of Josephus' redactional techniques to those of Luke and Matthew, see F. Gerald Downing, 'Redaction Criticism: Josephus' Antiquities and the Synoptic Gospels', *JSNT* 8 (1980), pp. 29–48; 9 (1980), pp. 46–65.; on bias and creative license in the Gospels compared to other biographical works, see: Michael Licona, 'Are the Gospels "Historically Reliable"? A Focused Comparison of Suetonius's Life of Augustus and the Gospel of Mark', Religions 10.3 (2019), 148; Michael Licona, *Why Are There Differences in the Gospels?* (New York: Oxford University Press, 2016); Craig Keener, 'Otho: A Targeted Comparison of Suetonius's Biography and Tacitus's History, with Implications for the Gospels' Historical Reliability', *BBR* 21.3 (2011), pp. 331–55.

[47] We did not perform an analysis on name origin statistics on the historical novels, because the origin data for many of these names was not available in Ilan-1.

Our research hypothesis assumes that Ilan was correct to generally classify the name occurrences from the GA and Josephus manuscripts as valid entries. Therefore, occurrences in GA and Josephus should 'fit' Ilan-1 but not fit Ilan-3; additionally, the three hypothetical samples (historical novels, Ilan-1F, and the uniform distribution) would likely not fit into any historical reference distributions. Considering our aims and the data discussed previously, this resulted in predictions concerning eighteen tests. See Table 2.

## 3. Results

### *3.1 Gospels & Acts, and Josephus*

The p-values of the eighteen chi-square goodness-of-fit tests are shown in Table 3. Their output tables, along with some discussion of the technical details may be found in the Supplementary Materials. For tests of GA and Josephus against Ilan-1, their names were removed from Ilan-1 in order to establish the independence of the two distributions, which is a condition for the chi-square goodness-of-fit test.

| Ref. Distr. | Variable | Gospels & Acts | Josephus | Novels | Ilan-1F | Uniform (ACT) |
|---|---|---|---|---|---|---|
| Ilan-1 | Name Freq. | 0.8556[48] | 0.0655[49] | $6.52 \times 10^{-15}$ | $2.20 \times 10^{-16}$ | $1.69 \times 10^{-15}$ |
| | Name Origin | 0.0034 | $1.09 \times 10^{-12}$ | NA | $2.20 \times 10^{-16}$ | $2.89 \times 10^{-6}$ |
| Ilan-3 | Name Freq. | $4.29 \times 10^{-13}$ | $2.20 \times 10^{-16}$ | $2.20 \times 10^{-16}$ | $2.20 \times 10^{-16}$ | $4.81 \times 10^{-5}$ |
| | Name Origin | $2.20 \times 10^{-16}$ | $2.20 \times 10^{-16}$ | NA | $2.20 \times 10^{-16}$ | 0.2819 |

*Table 3. Chi-square goodness-of-fit test results. Cell entries are p-values. The scientific notation, $1.19 \times 10^{-8}$, means the leading number (1.19 in this case) is divided by 10 to the 8th power. This means the decimal moves 8 places to the left, giving 0.0000000119 probability. The p-value $2.20 \times 10^{-16}$ is the smallest machine p-value without using extra precision. It is effectively zero. In the Uniform column, ACT stands for the GB "anonymous community transmission" hypothesis.*

Focusing on Table 3, we see that 16 out of the 18 tests matched our hypotheses from Table 2. Four of the p-values are greater than the Bonferroni-adjusted benchmark of *0.0028*, and the p-value for name frequencies in GA versus Ilan-1 is considerably high, providing evidence that GA fit Ilan-1 well. The two unexpected results are Josephus' name origin frequencies vs. Ilan-1 with p-value = $1.09 \times 10^{-12} < 0.0028$, which falls below our benchmark, and the name origin frequencies of ACT vs. Ilan-3 with a p-value *0.2819 > 0.0028*, which exceeds the benchmark.[50] The power for all tests was *0.999* or higher, so power values will not be further shown.[51]

[48] For this test, the p-value was 0.9217 if the name occurrences were not removed from Ilan-1.

[49] For this test, the p-value was 0.2143 if the name occurrences were not removed from Ilan-1. There were 274, or 12.5% of the names removed from Ilan-1 for the test in the table.

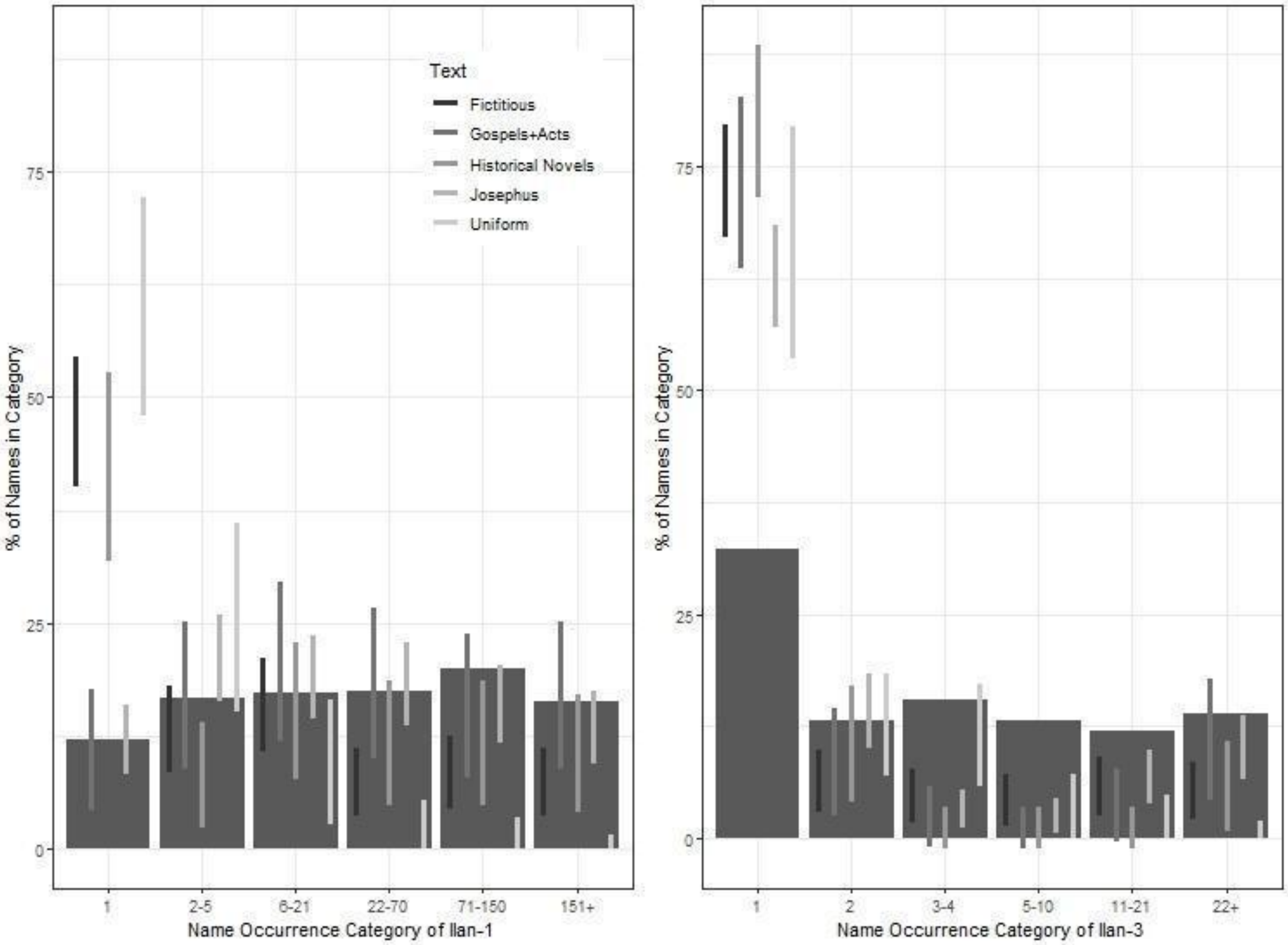

*Figure 2. Popularity Statistics Comparing All Test Distributions with Ilan-1 and Ilan-3. The bar is the reference distribution. The vertical lines are 95% confidence intervals. The match of Gospels & Acts, and Josephus, with Ilan-1 is seen with every confidence interval.*

In order to visualize the results of the hypothesis tests in Table 3, we have provided some figures which plot the reference distribution (Ilan) as a gray bar and the test distributions of GA, Josephus, the historical novels, Ilan-1F, and the Uniform distribution as grayscale 95% confidence intervals over the bar.[52] Figure 2 displays the popularity percentages and Figures 3 and 4 display the origin percentages. The way to interpret a single confidence interval is as follows: (i) the center of the interval is the actual proportion from the test distribution; (ii) the half-line above and below the center is the margin of error. If the interval goes through the top of the Ilan bar, then the test distribution is viewed as approximately fitting the reference for that single category. If the interval does not go through the top of the Ilan bar, then the test distribution is viewed as not fitting the reference for that single category. How close the interval is to the top of the bar matters; they can be 'close' and yet 'far off'. Keep in mind that the confidence intervals are for visualization, which is subjective. They are calculated independently of one another, and actual fit is determined by formal testing, which accounts for all categories and closeness of fit simultaneously.

[50] The statistical explanation is as follows: For Ilan-1, there are 457 unique names among the 2185 – meaning there are a lot of replicates (around 5 occurrences per name). By contrast, for Ilan-3, there are 575 unique names among the 1227 – meaning there are far fewer replicates (around 2 occurrences per name). This means that sampling from the 575 (uniform) is 'not too far' off from simply sampling from Ilan-3 itself, whereas sampling from the Ilan-1's 457 (uniform) is quite a bit different from sampling from Ilan-1's 2185.

[51] See Supplementary Materials for calculations.

[52] All confidence intervals are of the Wald type for proportions except the Uniform distribution, which had to be calculated using bootstrapping. For the Uniform bootstrap, a bootstrap sample came from one instance of each name in the reference distribution (457 names, each with probability 1/457), sampled 52 times without replacement (number of contested GA name occurrences). See Supplementary Materials for further details.

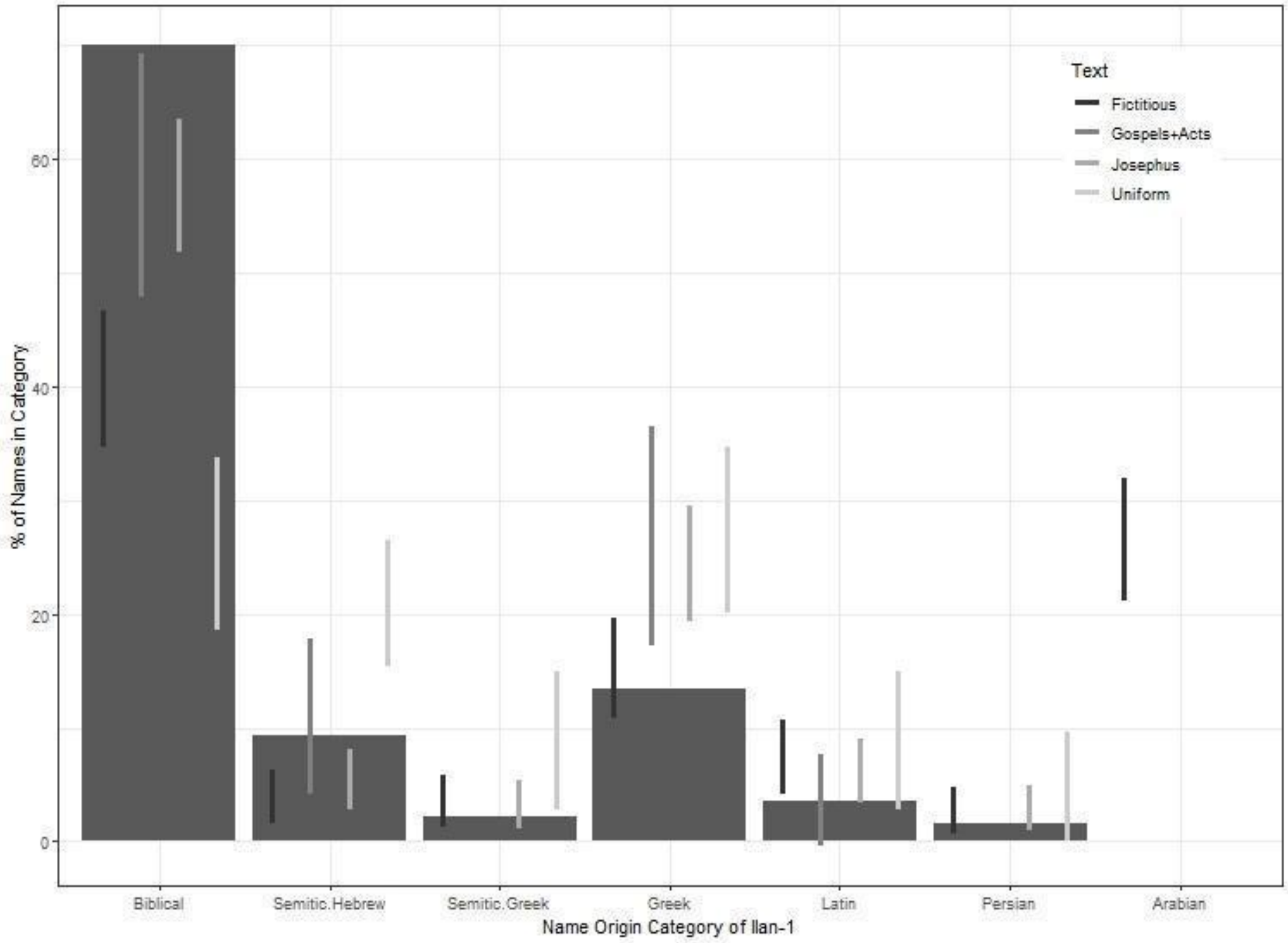

*Figure 3. Origin Statistics Comparing Gospels & Acts, Josephus, Ilan-1F, and Uniform with Ilan-1. The bar is the reference distribution of Ilan Vol. 1 Palestinian Males from 4 BCE to 73 CE. The vertical lines are 95% confidence intervals.*

Let us examine Figures 2, 3, and 4 and remark on their connection with the hypothesis tests from Table 3. In Figure 2, left side, we observe that all six intervals of GA and Josephus overlap Ilan-1, consistent with the high p-values in Table 3. GA is slightly low on occurrences of 71-150 times in Ilan-1, likely due to the fewer occurrences of Eleazar, while Josephus is high in occurrences of 2-5 and low in occurrences of 151+. Turning to the right side of Figure 2, only two of the six GA and Josephus intervals overlap Ilan-3 (2 and 22+), reinforcing the extremely low p-values in Table 3.

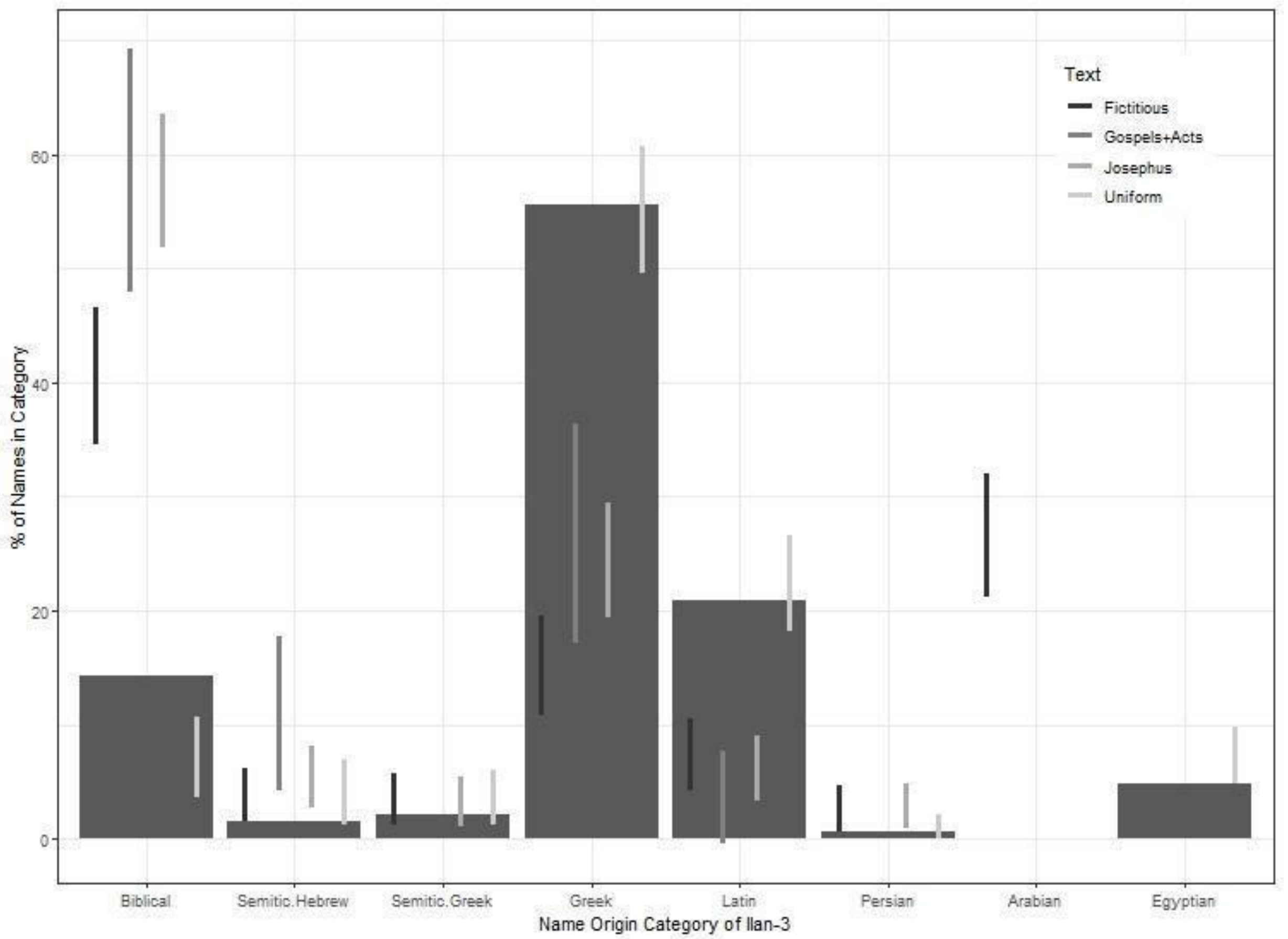

*Figure 4. Origin Statistics Comparing Gospels & Acts, Josephus, Ilan-1F, and Uniform with Ilan-3. The bar is the reference distribution of Ilan Vol. 3 Diaspora Males from 4 BCE to 73 CE. The vertical lines are 95% confidence intervals. The mismatch between Gospels & Acts, and Josephus, with Ilan-3 is visible across multiple categories.*

Moving to the origin statistics, we start with Figure 4 vs. Ilan-3. Again, neither GA nor Josephus visibly "fit." This is most prominent in that the biblical percentage of names is way too high whereas the Greek percentage is way too low. Semitic Hebrew names are also too high, balanced with Latin too low. Lastly, neither GA nor Josephus have any Egyptian names and therefore their percentage is 0 with no confidence interval able to be calculated. This is a clear case of no fit, as corroborated with the GA and Josephus p-values of $2.20 \times 10^{-16}$ in Table 3.

However, in Figure 3 things are not so clear. As previously, GA and Josephus track similarly. However, GA is slightly low on Biblical and high on Greek, but not much. Josephus is a little lower on Biblical and higher on Greek and Latin, and also a tad below on Semitic Hebrew. Notice also that the GA confidence intervals are always wider than Josephus. This is because Josephus' sample size is 274 whereas GA is only 82, providing less confidence and therefore wider intervals. All these factors combine to produce a figure that does not clearly show the difference in the magnitude of fit to the naked eye. But the p-value of the GA is 0.0034 whereas the Josephus p-value is $1.09 \times 10^{-12}$ in Table 3. This illustrates the reason why statisticians do not rely on subjective interpretations of graphs to draw their conclusions, but rather draw conclusions from appropriately chosen and applied statistical tests. One surprising result of our analysis, therefore, is that Josephus' name origin data statistically does not fit Ilan-1 due to a proportionately high occurrence of Greek and Latin names versus Biblical and Semitic Hebrew names.

Several factors likely contribute to this. Ilan's database relies on epigraphic as well as literary sources, including ossuaries from the region of Jerusalem, which show a preference for Biblical over Greek names. Conversely, Josephus' audience, focus, and literary milieu results in a greater preference for Greek over Biblical or Semitic names of Jewish persons. This is exemplified in his generally more Hellenized orthography and formal Greek case suffixes, while GA show a preference toward less official orthography coinciding, as Ilan notes, to common pronunciation.[53] Discussing the general prominence of Greek names in literary sources, Ilan observes:

> The literary texts record the lives of the more affluent and worldly, who chose names for their children from the culture by which they were influenced. The epigraphic record documents a wider variety of people and shows that this apparent trend in the literary sources was much less popular in the wider population.[54]

Compared to Josephus, GA provides a slightly more unvarnished, on-the-ground situation that more accurately reflects general naming practices, and the fitness of GA as opposed to Josephus in this instance is not merely due to the smaller sample size.[55]

### *3.2 Hypothetical Test Distributions*

Turning again to Figure 2, the occurrences from the historical novels and Ilan-1F vs. Ilan-1 generally do not fit, especially in the over-representation of infrequent names. This data runs counter to the argument from GB and is in accordance with observations made elsewhere that name rarity is typically over-represented by inventors of names.[56] Although we combined *Ben*

---

[53] Ilan, *Part I*, p. 18.

[54] Ilan, *Part I*, p. 40.

[55] The literary setting also influences the GA sample (e.g., 'Peter' instead of 'Cephas'), which is also tilted in favor of having higher Greek occurrences than expected due to the historical situation described in Acts 6. Note that adding the six Hellenists from Acts 6:5, as GB also observe ('Name Popularity', p. 185), significantly impacts the ratio of GA's Greek to biblical names in relation to the Ilan-1 reference distribution. Without the six Hellenists from Acts 6:5 (Stephen, Philip, Procorus, Nicanor, Timon, and Parmenas), the p-value of GA's origin statistics versus Ilan-1 is .0587, exceeding the standard threshold for statistical significance apart from the Bonferroni correction. The historical situation described in Acts readily supplies the reason why these Jewish males would have disproportionately borne Greek versus biblical names: they were specifically chosen to represent Hellenistic Jews (Acts 6:1–5). As an aside, combining Semitic-Hebrew and Semitic-Greek into a single category results in GA p-values of .065 (Hellenists excluded) and .0037 (Hellenists included). Were we to consider Cephas instead of Peter as reflecting the appropriate name origin of this occurrence (see John 1:42), the GA origin p-values would be 0.3915 (Hellenists excluded) and 0.0159 (Hellenists included). Bart Ehrman has suggested that Peter and Cephas were different persons ('Cephas and Peter', *JBL* 109 (1990), pp. 463–74), but the improbability of this position has been articulated by Dale Allison ('Peter and Cephas: One and the Same', *JBL* 111.3 (1992), pp. 489–95).

[56] See Van de Weghe, 'Name Recall', pp. 100–01.

*Hur* and *The Spear* into a single sample for our analysis, it is worth noting that their p-values fall significantly below our threshold when considered separately as well. *Ben Hur* has a p-value of $9.071 \times 10^{-6}$ and *The Spear* a p-value of $2.831 \times 10^{-15}$.

As for the uniform distribution vs. Ilan-1, there is an over-representation in frequency categories '1', but severe under-representation in '71-150' and '151+'. This is not surprising since each name was represented only once. As the p-values in Table 3 demonstrate, testing GA and these hypothetical samples against Ilan-1 demonstrates that at least for a male Jewish Palestinian population (4 BCE–73 CE), the number of name occurrences in GA has been shown to fit the reference distribution while these other test distributions failed to fit.

Let us now briefly consider the 26 uncontested GA name occurrences as cataloged by GB.[57] When considering these against a reference distribution of Ilan-1, they have a p-value of *0.1369* which is well above the *0.0028* benchmark (cf. contested names only *p-value = .7450*, complete GA sample *p-value = 0.8543*). Adding the contested names so as to consider the full sample of GA names results in at least as good of a historically situated fit. This would be expected if the contested name occurrences are authentic.[58]

Even if, contrary to the evidence we possess,[59] accurate and intricate naming patterns were achieved by a fictionalizing author of antiquity, there are several additional reasons to doubt that the name statistics in GA result from a fictionalizing process. First, in second- to third-century Gospels like *The Sophia of Jesus Christ*, *The Gospel of Judas*, and *The Gospel of Mary*, any hint of Jesus' Palestinian environment is overshadowed by mythological concerns and persons.[60] Second, Gospels like *The Infancy Gospel of Thomas* and *The Infancy Gospel of James* may add an unusual name like Zenon, a New Testament name like Zaccheaus, or a popular Western Diaspora name like Samuel, but the general trend reflects a disinterest in capturing any kind of realistic onomastic verisimilitude.[61] This is seen, particularly, in the lack of qualifiers to disambiguate or distinguish persons with popular names in compositions like *The Gospel of Mary* and *The Coptic Gospel of Philip* (for more on name disambiguation, see below).[62] Third, to

---

57 Gregor and Blais, 'Name Popularity', pp. 185–86.

58 We thank Willem Jan Blom for bringing this point to our attention.

59 Van de Weghe, 'Name Recall', pp. 101–04.

60 *The Sophia of Jesus Christ* incorporates the name Yaldabaoth but replaces the name 'Jesus' with the terms 'Savior' and 'Lord' (See Simon Gathercole, *The Apocryphal Gospels* (Westminster, London: Penguin Books, 2021), p. 304); *The Gospel of Judas* includes the names Barbelo, Autogenes, Adamas, El, Nimrod-Yaldabaoth, Christ, Harmathoth, Galila, Yobel, and Adonaios (Gathercole, *The Apocryphal Gospels*, p. 194); *The Gospel of Mary* incorporates personifications of Desire and Ignorance and also replaces the name 'Jesus' with the term 'Savior' (Gathercole, *The Apocryphal Gospels*, p. 254).

61 For a survey of named characters in apocryphal Gospels, see the lists of *dramatis personae* in Gathercole, *The Apocryphal Gospels*, pp. 4, 31, 48, 147–48, 155, 163, 194, 207–11, 254, 261, 304, 361. *The Infancy Gospel of James* contains 12 Jewish male name occurrences, including two Samuels (cf. Gathercole, *The Apocryphal Gospels*, p. 4; note that Gathercole's lists are not exhaustive); *The Infancy Gospel of Thomas* contains six Jewish male name occurrences, including Thomas 'the Israelite', an inappropriate qualifier (see Gathercole, *The Apocryphal Gospels*, p. 31; cf. Ilan, *Part 1*, pp. 32–36). On the comparative scarcity of onomastic data in various apocryphal works and their incongruity with Ilan-1, see Van de Weghe, 'Name Recall', pp. 100–01.

62 The lack of appropriate qualifiers in *The Infancy Gospel of Thomas*, *The Infancy Gospel of James*, *The Gospel of Peter*, and *The Gospel of Mary* is discussed in Van de Weghe, 'Name Recall', p. 100. *The Gospel of Mary,* for assume that a later author would diligently examine a source like Josephus or other material from Palestine, take a rough inventory of name popularity, and attempt to reproduce it assumes, against available evidence, a hyper-sensitive focus on names that no critical NT reader considered meaningful for two millennia until Richard Bauckham began the research that resulted in *Jesus and the Eyewitnesses*. These three observations align with the published data on personal name retention, which suggests that names, being arbitrary and easily forgettable, are among the least integrated and lasting pieces of information in the recollection of stories or experiences.[63]

In this regard, we must also consider the unique historical setting in which our GA sample occurs. While naming frequencies in the Greco-Roman world were generally distributed widely and evenly, the uniquely high occurrences of only several Hasmonean names make it much harder for a fictional author to achieve a level of apparent authenticity.[64] This explains why, surprisingly, the name origin frequencies of 'anonymous community transmission' (uniform) vs. Ilan-3 show a fit with a p-value of 0.2819 with a sample size of 52 (GA contested name only size) and 0.2026 with a sample size of 82 (full GA size), which indisputably exceeds the benchmark.[65] This demonstrates that while it is possible that an 'anonymous community transmission' (uniform) distribution may fit some real reference distributions (Ilan-3), it does not fit others (Ilan-1). This provides evidence that the fit signal from GA to Ilan-1 can be statistically distinguished from the lack of fit signal of GA to Ilan-3.

This situation in Ilan-1, wherein a relatively high number of persons bear relatively fewer names, creates further patterns of historicity in GA that are more difficult to quantify. Note, for example, that from the 28 cases in which a GA author disambiguates a person's primary name (e.g., Simon *the Leper*, Judas *Iscariot*, etc.), 64% of these belong to persons bearing one of the top five names and 82% to persons bearing one of the top 12 names.[66] This reflects, again, a 'situation on the ground' wherein men bearing the most common names would naturally need to

---

example, never specifies which Mary is being referred to (although Mary Magdalene is most probable); *The Coptic Gospel of Philip* contains three Marys, but only one Mary is qualified (see Gathercole, *The Apocryphal Gospels*, p. 254, 361).

[63] See the discussion in Van de Weghe, 'Name Recall,' pp. 95–96.

[64] As Ilan observes, 20.7% of the name pool served 73.4% of the male population (*Part I*, p. 5). To illustrate the uniqueness of this, consider the top two names in Ilan-1. These belong to more than 16% of males. The top two Greek names in Coastal Asia Minor belong to under 5% of the male population (1,849 out of 39,477 occurrences; P. M. Fraser and E. Matthews, eds., *The Lexicon of Greek Personal Names, Volume V.B, Coastal Asia Minor: Caria to Cilicia* (Oxford: Clarendon, 2013), pp. xxx-xxxi).

[65] Part of the reason is due to the uniformity of name occurrences within the Ilan-3 population compared to Ilan-1. For Ilan-1, there are 457 unique names among 2185 occurrences, meaning there are more replicates. This averages to about five occurrences per name, but in reality some occur with high frequency while many are rare (only one occurrence). By contrast, for Ilan-3 there are 575 unique names among 1227 occurrences, meaning there is far less high/low frequency variation. That fact alone does not result in a fit with uniform distribution frequencies (p-value = 2.20×10-16), but that combined with the particular mix of frequency-origin combinations can.

[66] The number 28 excludes qualifiers due to titles (e.g., Caiaphas the high priest, king Herod, etc.): Simon (6), Joseph (4), Jacob (4), Yohanan (3), Judah (4), Joshua (2), Mattathias (1), Andreas (1), Eleazar (1), Levi (1), and Nathanel (1). Including titles brings the total number of qualified name occurrences to 36, adding the following: Qaifa (1), Hanan (1), Agrippa (1), Philip (1), Hananiah (1), and Herod (3). Even with this number, 75% of qualifiers are given to names ranking in the top 12 of Bauckham's table 6 in *Jesus and the Eyewitnesses*, p. 84.

be distinguished from one another; it is contrary to the practice in De Wohl's *The Spear*, which shows a tendency toward qualifying names regardless of their popularity.[67] In GA, the situation becomes even more precise. For example, Lazarus 'of Bethany' is disambiguated (John 11:1); this name occurs widely in the general population, but not in GA itself. The common name 'Jesus' is consistently disambiguated throughout the Gospels in the public speech of characters but not by the Gospel authors themselves in segments of narration, as would be expected under an authentic information-retention scenario.[68]

Much like the less official orthography of GA, the naming practices at a meticulous level favor a model that allows for natural name retention on a considerable scale. The data does not suggest that the name popularity distribution in GA results from the achievements of well-informed inventors, let alone that it is a product of 'anonymous community transmission', which was 'freely created and modified according to the needs of the community…communities that had no interest in the past and had no reason to attempt to preserve historical accounts for their historical value.'[69] What can account for the significant, accurate name retention in GA? Rather than pointing toward a model driven by creative or uncontrolled development, it suggests a model that aligns with current understandings of ancient historiographic practices, especially for history-writing of contemporary or recent events; such a model involves dependence on primary source materials, data compilation, inquiry of participants, and, in line with Bauckham's broader study, a preference for eyewitness testimony.[70] In short, the evidence supports our research

---

[67] Of the 25 qualified names in *The Spear*, over half have under ten occurrences in Ilan-1 (all besides Nathan and Hillel have five or less): Boz, Zadok, Aaron, Ephraim, Achim, Oziah, Amram, Nathan, Mordecai, Josaphat, Nicodemus, Baruch, and Hillel. When *The Spear* includes a qualifier with a popular first name, it is often because the entire name, including qualifier, is from GA (e.g., Simon bar Jonah, Judah bar Alphaeus, Yohanan bar Zebedee, Annas the high priest, Joseph of Arimathea, Jesus bar Joseph, John the Baptist, Judas from Kerioth). We thank Lydia McGrew for drawing our attention to this trend.

[68] This is because the authors clearly knew which Jesus they were talking about, while in the general public his name would require a disambiguation. Peter Williams details this phenomenon in *Can We Trust the Gospels?* (Wheaton, IL: Crossway, 2018), pp. 71–75.

[69] Bauckham, *Jesus and the Eyewitnesses*, p. 245. In this passage he is primarily referring to the form criticism of Rudolph Bultmann, *The History of the Synoptic Tradition* (tr. J. Marsh; Oxford: Blackwell, rev. ed. 1968), but also to the pioneering works of Martin Dibelius, *From Tradition to Gospel* (tr. B. L. Woolf; London: Nicholson and Watson, 1934) and Karl L. Schmidt, *Der Rahmen der Geschichte Jesu* (Berlin: Trowizsch, 1919) whose scholarship Bauckham has in mind when using the words 'anonymous community transmission' (see Bauckham, *Jesus and the Eyewitnesses*, pp. 7, 8, 294, and 304; cf. 245).

[70] For this preference among ancient historians, see: J. J. Peters, *Luke Among the Ancient Historians* (Eugene, OR: Pickwick, 2022); Samuel Byrskog, *Story as History, History as Story: The Gospel Tradition in the Context of Ancient Oral History* (Tübingen: Mohr Siebeck: 2000); pp. 49–91; Guido Schepens, 'History and *Historia*: Inquiry in the Greek Historians', in *A Companion to Greek and Roman Historiography* (ed. John Marincola; Oxford: Blackwell, 2007), pp. 39–55; John Marincola, *Authority and Tradition in Ancient Historiography* (Cambridge: Cambridge University Press, 1997). The high name retention in GA also aligns with a model that incorporates a form of controlled versus uncontrolled oral tradition (see Eric Eve, *Behind the Gospels: Understanding the Oral Tradition* (London: SPCK, 2013); Kenneth Bailey, 'Informal Controlled Oral Tradition', *Themelios* 20 (1995), pp. 4–11 (7–8); cf. Theodore Weeden, 'Kenneth Bailey's Theory of Oral Tradition: A Theory Contested by Its Evidence', *JSHJ* 7.1 (2009), pp. 3–43; James Dunn, 'Kenneth Bailey's Theory of Oral Tradition: Critiquing Theodore Weeden's Critique', *JSHJ* 7.1 (2009), pp. 44–62; Rainer Riesner, *Rainer, Jesus als Lehrer: Eine Untersuchung zum Ursprung der Evangelien-Überlieferung* (Tübingen: J. C. B. Mohr: 2nd ed., 1984); Birger Gerhardsson, M*emory and Manuscript: Oral Tradition and Written Transmission in Rabbinic Judaism and Early*

hypothesis. GB's contention that 'Bauckham's thesis offers no advantage in explaining the observed correspondence between name popularity in Gospels-Acts and in the contemporary Palestinian Jewish population over an alternative model of "anonymous community transmission"' conflicts with the available evidence.[71]

## 4. Remarks on Gregor and Blais

Having completed our analysis of the data, we now explicitly address the work of GB in the following remarks. These highlight why our findings were distinct from GB and where GB's statistical analysis falls short.

(4.1) The primary argument of GB rests on their Figures 1 and 2 which show 95% confidence intervals of name frequencies, somewhat like our own Figures 2, 3, and 4. The first major problem with this approach is that their conclusion does not necessarily follow from their graphs. We disagree with their interpretation and would actually draw the opposite conclusion from their graphs.[72]

(4.2) The second major problem with GB basing their argument on their Figures 1 and 2 is that they implicitly assert a statistical hypothesis test[73] with a null hypothesis of 'anonymous

*Christianity with Tradition and Transmission in Early Christianity* (Biblical Resource Series; Grand Rapids, MI: Eerdmans 1998); Tuomas Havukainen, 'Birger Gerhardsson on the Transmission of Jesus Traditions – How Did the Rabbinic Model Advance a Scholarly Discourse?', *Iesus Aboensis* 1 (2015), pp. 49–63. For other discussions relating name retention to historiography, see Byrskog, *Story as History*, pp. 266–306; Bauckham, *Jesus and the Eyewitnesses*, pp. 524–35; Simon Hornblower, 'Personal Names and the Study of the Ancient Greek Historians', in *Greek Personal Names: Their Value as Evidence* (eds. Simon Hornblower and Elaine Matthew; Oxford: Oxford University Press, 2000), pp. 129–43 (139–40); R. Shroud, 'Thucydides and Corinth', *Chiron* 24 (1994), pp. 267–302. The explanatory power of these considerations to account for the high name retention in GA is discussed at length in Van de Weghe, 'Name Recall', pp. 105–09.

[71] Gregor and Blais, 'Name Popularity', p. 171.

[72] GB write, 'Most observed Gospels-Acts numbers of name occurrences fit inside the confidence interval…' (p. 24). In fact, not merely 'most', but 25 out of the 26 names shown fit GB's confidence intervals generated from Ilan-1. Furthermore, 7 out of 26 names shown do not match their uniform distribution. We contend that this shows better fit to Ilan-1 than Uniform. This interpretive disagreement is the reason for employing the chi-square goodness-of-fit test from Section 3.

[73] Statistical hypothesis tests consist of two hypotheses: the null hypothesis and the alternative hypothesis. They are structured such that the null hypothesis is the status quo and assumed to be true unless the preponderance of evidence from the data indicates otherwise, in which case the alternative hypothesis is concluded. In our paper, 'the preponderance of evidence' is given by the benchmark of 0.05, divided by the 18 tests, which is 0.0028. This is an objective criterion: when a p-value is above 0.0028 the data supports the null hypothesis, when the p-value falls below 0.0028 the null hypothesis should be rejected in favor of the alternative, in the absence of countervailing evidence.

community transmission', or random name selection.[74] The problem is that they went contrary to standard statistical practice and reversed their null and alternative hypotheses.[75]

(4.3) Third, using the method of GB, a sample of size 53 (or 52) from many other text's distributions would draw the same conclusion of 'fit' to the uniform distribution, i.e. 'anonymous community transmission'. To see this, look at their Figure 1: the top of the gray bar passes through, or at least touches, all of the confidence intervals shown. Because of this, GB argue that GA fits the uniform distribution. By contrast, in their Figure 2, not all of the Josephus confidence intervals go through the gray bar, therefore Josephus does not fit the uniform distribution, they argue. The problem is that if Josephus' sample were size 53, then the confidence intervals of Figure 2 would correspondingly widen and look like Figure 1, switching the conclusion of 'no fit' to 'fit'. No one thinks Josephus' writings were 'not statistically significantly different'[76] from anonymous community transmission – yet GB's methodology would conclude this if he had merely written less! This is a flaw in their methodology.[77] This fallacy arises from switching the null and alternative hypotheses as in the above remark.

This problem is accentuated by the high power for these goodness-of-fit tests. We have shown in Section 3 that the power is *0.999+*. This means that, for cases like those under consideration (6+ bins, many names, and sample sizes 50+), when a null hypothesis is rejected, we can be confident that the test distribution does not fit the reference. How to interpret the lack of fit needs to weigh all the evidence together. For example, even though we accept that

---

[74] GB never explicitly state a formal statistical hypothesis test. However, some (not all) of their language, approach, and Supplementary Materials refer to an implied hypothesis test. This is most clearly seen with their use of the technical phrase 'statistically significantly different' on p. 197 of 'Name Popularity', which refers to rejecting a null hypothesis in favor of the alternative hypothesis:

> Any combination of at least some of the contested characters being historical and the invention of fictitious characters' names with at least some information about name popularity would result in a name popularity distribution corresponding to the contemporary population distribution more closely than the distribution generated on the most extreme scenario. Because the sample size of Gospels-Acts name occurrences is so small that the observed Gospels-Acts name popularity distribution is already not *statistically significantly different* from [anonymous community transmission]" (emphasis ours).

[75] In statistical goodness-of-fit testing, of which the chi-square goodness-of-fit test is the appropriate test for our data type, the null hypothesis is 'the data fit the reference distribution'. See the dice example in Section 2.1. In our case, the reference distribution must be the historical Ilan, or something of that sort. This is simply how the tests work, and GB have silently switched their null and alternative hypotheses. The reason the null must be the reference distribution is that it is well-defined, whereas the possible alternative distributions are uncountable. They would certainly include 'anonymous community transmission', but which one? Defining this 'anonymous community transmission' distribution is problematic. See Supplementary Materials for some discussion. GB opt for a uniform distribution based on the 451 unique names in their adjusted Ilan-1 plus the three names they returned to Ilan-1. Why not use a mixture uniform distribution with two classes from that set? Why not use the set of just contested/uncontested New Testament names? The point is that there are an uncountable number of alternative possibilities and that is precisely the reason that goodness-of-fit testing places the reference distribution in the null hypothesis. This is why there are different values of power, depending on which alternative distribution is used; see the dice example in Section 2.1.

[76] Gregor and Blais, 'Name Popularity', p. 197.

[77] Additionally, Ilan-1F demonstrates that forgers often incorporated names into their narratives that occur nowhere in Ilan-1; their consideration of only Palestinian names is not justified by the small sample of occurrences from the *Infancy Gospel of James*. Regardless, the p-value of 52 uniform ACT occurrences vs. Ilan-1 is significantly lower ($1.69\times10^{-15}$) than the p-value of the 52 contested GA occurrences vs. Ilan-1 (0.7450).

Josephus' names do not match Ilan-1's origin distribution, we do not therefore conclude that the sample is not historical, rather we see a historical reason for his over-representation of Greek names (see discussion in Section 3.1).

(4.4) Fourth, the confidence intervals employed by GB in Figures 1 and 2, as well as in our own Figures 2, 3, and 4, assume independence between names, which is false. Such confidence intervals are used because they are easy to calculate, relatively easy to explain, and make a great picture to illustrate the point. Conclusions using the confidence intervals would be valid for only the single name (GB Figures 1 and 2) or category (our Figures 2, 3, and 4) in reference; using more than one ignores the multiple testing problem described in Section 2.1. Thus, the proper use of the figures is for illustration only. In reality, the proportion of one name is dependent upon the proportions of all of the other names, which is not accounted for in the confidence interval calculations. However, this dependence relationship is precisely accounted for in the chi-squared goodness-of-fit test we have used earlier in this paper.

(4.5) Fifth, GB assert that names occurring only once in GA provide no information about whether GA fits the historical distribution.[78] However, as we have shown above, grouping individuals in order to meet the test conditions is a common issue in goodness-of-fit testing, and the information for every name can be accounted for by binning them into appropriate groups, as we have done.

(4.6) Sixth, GB's Section 5 directly addresses the issue of rare names. This is a welcome step. Rare names are those which occur only once in Ilan-1.[79] We overlook the fact that GB previously claimed that names occurring only once in GA were 'without any information about name popularity'.[80] However, the method used in this section has substantial problems. To start, they reversed the null and alternative hypotheses which they used in Section 4. Although we agree that this is now the correct configuration of the hypotheses, it is an outright inconsistency.[81]

The main issue is that GB's conclusion in Section 5 is highly sensitive to the number of rare names in the list, and we dispute their list, which stands at four (Aeneas, Agabus, Bartimaeus, and Timaeus). In their handling of the data, GB removed 26 'attested'[82] name

---

[78] 'Therefore, most of the Gospels-Acts data is entirely uninformative and cannot support Bauckham's thesis. It is statistically indistinguishable from "white noise" generated entirely randomly without any information about name popularity' (Gregor and Blais, 'Name Popularity', p. 191).

[79] In their article, GB define rare names as 'attested only once among first-century Palestinian Jews' ('Name Popularity', pp. 174, 198). If any readers wish to use GB's Ilan file, as we have done, note that it defines rare as either attested once (262 names), or twice (79 names, except "Yeshab" appears to have inadvertently not been marked as rare).

[80] See the fifth remark in this section.

[81] See our remarks in 4.2 and 4.3. In GB's Section 4 their null hypothesis was anonymous community transmission and the alternative was Ilan-1. Now, in GB's Section 5 their null hypothesis is Ilan-1 and the alternative is anonymous community transmission.

[82] Removing the 26 'attested' names from the GA sample is built on questionable assumptions that we do not agree with. Foremost, it alters the GA dataset to favor their conclusion (in this case, removing names from a dataset and then concluding that the dataset is too small to determine statistical significance). Further, removing 'attested' names is justified by a narrow focus on several overstatements made by Bauckham (Gregor and Blais, 'Name Popularity',

occurrences (21 names) from Bauckham's original 79 (45 names) to obtain their 53 (32 names) unattested occurrences.[83] The names removed include two rare names (Qaifa and Toma). According to their calculations, the 'probability of getting [four or fewer] names … is only 1.1%!'[84] According to GB's argument, the table below shows the number of rare names and the probability of that many or fewer occurring if the Ilan-1 distribution was true.[85] It can be seen that their conclusion is extremely sensitive to the number of rare names. GB's list gives a 1.1% probability, which would jump to 7.3% if just the two names were not excluded (Table 4). Moreover, if the Palestinian Acts Hellenists were added, as we believe they should be, it includes four more 'rare' names and would bring the list to eight or ten rare names at 24% or 50% of occurring (Table 4), depending on whether we also add the two contested rare names, Qaifa and Toma.[86]

| **# rare** | **3** | **4** | **5** | **6** | **7** | **8** | **9** | **10** |
|---|---|---|---|---|---|---|---|---|
| % | 0.3 | 1.1 | 3.2 | 7.3 | 14 | 24 | 36 | 50 |

*Table 4. Probability of obtaining numbers of rare names. This table is based off GB's Figure 3. For example, # rare = 6 and %=7.3 means that there is a 7.3% probability that 6 or fewer rare names would be obtained in a sample of size 53 selected randomly from GB's pool of 2,582 occurrences.*

## 5. Conclusion

In conclusion, GB's analysis suffers from a lack of proper statistical methodology, thereby introducing critical errors into their analysis. In this paper, we have attempted to correct their missteps and to analyze as much data as possible with the best available statistical methods. We have shown how chi-square goodness-of-fit testing establishes that naming patterns in the Gospels and Acts fit into their historical context well, and that they fit statistically significantly better than fictitious works from antiquity and well-researched historical novels from the modern era. We have considered the Gospels and Acts name sample in light of two reference distributions and also discussed name origin data, sample size, and less quantifiable observations

---

p. 179, 180, 181; see also, Van de Weghe, *Living Footnotes*, p. 34), while a broader reading of Bauckham could allow all the GA data to be considered as part of an ongoing scholarly discussion regarding the historicity of GA. Further, the notion of 'attested' is based on ad hoc considerations; GB do not try to argue that GA authors relied on any of these sources for their name statistics, for example, but merely that we can know from other sources that these names belonged to historical persons. This creates a very uncomfortable scenario in which historically-attested GA names function to *undermine* the historicity of names in GA within GB's argument.

[83] The reason the 45 names only reduce to 32 names when 21 names are removed is that 8 names have separate occurrences in both lists.

[84] Gregor and Blais, 'Name Popularity', p. 198.

[85] See GB's Figure 3. The actual probabilities are not shown in the article, but they are closely approximated by a binomial distribution with n=53 and p=0.53, which are the results shown, including the 1.1%. See Supplementary Materials.

[86] Since we remove Bartimaeus, our sample includes nine rare GA names. Reflecting Ilan's orthography, they are: Hagaba, Timaeus, Qaifa, Parmenas, Procharus, Stephanus, Toma, Timon, and Aeneas. Aeneas is attested once in Ilan-1 outside of GA, so including this in the rare category depends on how one categorizes the GA occurrence. If the occurrence 'Peter' is cataloged under the name 'Cephas', an additional rare name would be added.

about name disambiguation and naming practices within invented materials. The evidence suggests that GA accurately retained personal names – those unmemorable pieces of personal information – at a remarkably high level, reflecting characteristics more in line with Josephus than with ahistorical and novelistic samples. In the words of Richard Bauckham, 'All the evidence indicates the general authenticity of the personal names in the Gospels.'[87]

[87] Bauckham, *Jesus and the Eyewitnesses*, p. 84.